\documentclass[aps,twocolumn,noshowpacs]{revtex4}

\usepackage{graphicx}  
\usepackage{amsmath,amssymb}   

\usepackage{tensor}	
\newcommand{\ind}[1]{\indices{#1}}

\usepackage{float,multirow}	

\usepackage{hyperref}	
\hypersetup{
	colorlinks,
	citecolor=blue,
	linkcolor=blue,
	urlcolor=blue,
	linktocpage=true
}
\begin{document}

\title{The Weak Gravity Conjecture, RG Flows, and Supersymmetry}

\author{Anthony M. Charles} 
\affiliation{Institute for Theoretical Physics, KU Leuven, Celestijnenlaan 200D, B-3001 Leuven, Belgium}

\begin{abstract} 
\noindent We study one-loop divergences in Einstein-Maxwell theory and their implications for the weak gravity conjecture.  In particular, we show that renormalization of these divergences leads to positivity of higher-derivative corrections to the charge-to-mass ratio of dyonic black holes.  This allows charged extremal black holes to decay into smaller ones, and so the weak gravity conjecture is automatically satisfied.  We also extend this analysis to a much wider class of Einstein-Maxwell theories coupled to additional massless matter fields and find the same result.  We then go on to study one-loop divergences in $\mathcal{N} \geq 2$ supergravity and show that dyonic black holes in these theories are protected against one-loop quantum corrections, even if the black hole breaks supersymmetry.  In particular, extremal dyonic black holes are stabilized by supersymmetry and cannot decay.

\end{abstract}

\maketitle

\section{Introduction}

The weak gravity conjecture (WGC), in its simplest form, posits that any quantum field theory with a $U(1)$ gauge symmetry that can be UV-completed into a theory of gravity must contain a state that is \emph{super-extremal}, i.e. has a charge-to-mass ratio that exceeds one.  The original motivation for this was to provide a way for extremal charged black holes (whose charge-to-mass ratio is exactly one) to decay, in order to prevent a huge number of black hole remnants from piling up throughout the universe~\cite{ArkaniHamed:2006dz}.  Since then, the WGC and its fellow other swampland conjectures (see e.g.~\cite{Ooguri:2006in,Brennan:2017rbf,Ooguri:2018wrx,Reece:2018zvv,Harlow:2018tng}) have been proposed as ways to delineate the properties that an effective field theory (EFT) must satisfy in order to have a UV-completion into a theory of quantum gravity.

One approach for understanding the WGC comes from~\cite{Kats:2006xp}, where it was shown that, in four-dimensional Einstein-Maxwell theory without a cosmological constant, the extremality condition for charged black holes is modified by higher-derivative corrections.  This modification is controlled by a particular combination of coefficients of the higher-derivative operators in the action, which we will denote by $\Delta$.  If this combination is positive, then super-extremal black holes are allowed, and thus the WGC is satisfied.  This condition has been argued for in a number of different ways, including monotonicity of the Wald entropy~\cite{Cheung:2018cwt,Cheung:2019cwi}, positivity of scattering amplitudes~\cite{Bellazzini:2019xts}, and field-theoretic sum rules that arise when the UV can be described by a CFT~\cite{Mirbabayi:2019iae}.

The goal of this paper is to provide a much simpler way to show that $\Delta > 0$: when one-loop quantum corrections to our theory are accounted for, the effective field theory parameters become coupling constants with an associated renormalization group (RG) flow.  This RG flow will guarantee that $\Delta$ will always become positive in the deep IR of our theory, regardless off what value it takes at higher energies.  The deep IR region of our quantum theory is precisely the region relevant for very massive black holes, and so the WGC will be automatic.

This RG-based analysis of the WGC is useful for numerous reasons.  It is a conceptually simple approach that does not rely on any assumptions on the actual UV dynamics of the gravitational theory; the only requirement is that at sufficiently low energies, the theory can be described by an effective field theory with a $U(1)$ gauge field (i.e. a photon) and a graviton.  Moreover, the techniques used to compute the running of EFT parameters are easily applied to more general low-energy field content.  In particular, we show that minimally-coupled fields with spin $s \leq 1$ can only enhance the running and make the super-extremality shift larger.  However, by including fields of spin $s = 3/2$ fields into the EFT, or by including fields with non-minimal couplings to the background gauge field, it is possible to change the sign of the running such that super-extremal black holes are forbidden.

This points us very naturally to consider supersymmetric settings that require gravitino fields.  In particular, we show that dyonic black holes in $\mathcal{N}\geq 2$ supergravity are not allowed to be super-extremal; the massless field content dictated by supersymmetry stabilizes the RG flow and keeps the parameter $\Delta$ locked at zero.  In this way, we recover the BPS bound on black holes entirely with EFT techniques.  Moreover, we show that there are no corrections to dyonic black holes from four-derivative operators in the EFT, even when one-loop quantum effects are accounted for.

Throughout this paper, we will focus on single-centered dyonic Reissner-Nordstr\"om black holes.  Our analysis can be extended to more general black holes with additional matter fields turned on in the background, including multi-centered black holes, but the higher-derivative corrections in such settings are more difficult to analyze~\cite{Natsuume:1994hd}.  Moreover, it has been proposed in~\cite{Palti:2017elp} that the WGC itself is modified in the presence of non-trivial scalar field profiles, and so one would have to carefully reconsider whether higher-derivative corrections allow for black holes themselves to satisfy this more elaborate form of the WGC.

The organization of this paper is as follows.  In section~\ref{sec:em}, we review higher-derivative corrections to Einstein-Maxwell theory and show how they shift the extremality bound for dyonic black holes.  In section~\ref{sec:qm}, we compute quantum corrections to Einstein-Maxwell theory and show how they lead to RG flows for the EFT parameters, which in turn allows for super-extremal black holes, even in the presence of additional minimally-coupled matter fields.  In section~\ref{sec:n2} we study quantum corrections to black holes in $\mathcal{N}\geq 2$ supergravity to show that super-extremal black holes are forbidden (thus preserving the BPS bound), before ending with concluding remarks in section~\ref{sec:con}.

\section{\label{sec:em} Einstein-Maxwell Theory}
Let's consider a four-dimensional low-energy EFT, valid up to some UV cut-off scale $\Lambda$, where the only massless fields are a photon and a graviton.  We will also assume that the spacetime has no cosmological constant.  As is standard for black hole applications, we will organize operators in the EFT according to the number of derivatives that appear in each operator.  At leading order (i.e. at two-derivative order), the Lagrangian for our theory is simply the Einstein-Maxwell Lagrangian
\begin{equation}
	\mathcal{L} = \frac{1}{2\kappa^2} R - \frac{1}{4\kappa^2} F_{\mu\nu}F^{\mu\nu}~,
\label{eq:l2}
\end{equation}
where $\kappa^2 = 8\pi G = M_\text{pl}^{-2}$.  In the spirit of effective field theory, we should also include all higher-order operators that are consistent with the symmetries of our theory.  In the absence of any charged sources, there are na\"ively eight independent operators we can write down at four-derivative order that are parity-invariant~\cite{Kats:2006xp}.  However, we are free to make field redefinitions of the form $g_{\mu\nu} \to g_{\mu\nu} + \delta g_{\mu\nu}$, as all physical quantities should be independent of such redefinitions.  In particular, we will choose a field redefinition that imposes the ordinary Einstein-Maxwell equations of motion on the higher-derivative operators.  This allows us to, without loss of generality, choose a basis of higher-derivative operators of the form
\begin{equation}\begin{aligned}
	\Delta \mathcal{L} &= c_1 W_{\mu\nu\rho\sigma}W^{\mu\nu\rho\sigma} + c_2 E_4 \\
	&\quad+ c_3 R_{\mu\nu\rho\sigma}F^{\mu\nu}F^{\rho\sigma} + c_4 (F_{\mu\nu}F^{\mu\nu})^2~,
\label{eq:l4}
\end{aligned}\end{equation}
where $W_{\mu\nu\rho\sigma}$ is the Weyl tensor, $E_4 = R_{\mu\nu\rho\sigma}R^{\mu\nu\rho\sigma} - 4 R_{\mu\nu}R^{\mu\nu} + R^2$ denotes the four-dimensional Gauss-Bonnet invariant and $c_i$ are some dimensionless parameters.  We want to stress that these Wilsonian parameters are not fixed; as we will show in the proceeding section, they should be thought of as coupling constants that have an associated RG flow.

We have so far only included relevant and marginal operators into our EFT, but we can in principle go further and also include irrelevant operators; their contribution to the Lagrangian will take the schematic form
\begin{equation}
	\Delta \mathcal{L} = \sum_n \frac{\lambda_n}{\Lambda^{n-4}} \mathcal{O}_n~,
\end{equation}
where $\lambda_n$ are dimensionless coupling constants, $\Lambda$ is the cut-off scale, and $\mathcal{O}_n$ is an $n$-derivative operator, e.g. $R^{n/2}$, $R^{n/2-1} F^2$, etc.  These operators are therefore suppressed by powers of the cut-off scale, and so their contribution to physical observables will be subleading.

Let's now consider a dyonic black hole of mass $M$, electric charge $Q$, and magnetic charge $P$ as a solution to the equations of motion of our leading-order Einstein-Maxwell Lagrangian (\ref{eq:l2}).  If we define $m = \kappa^2 M / 8\pi$, $q = \kappa Q/4\sqrt{2}\pi$, and $p = \kappa P/4\sqrt{2}\pi$ as the mass and charges in units of the Planck length, the metric takes the form
\begin{equation}\begin{aligned}
	ds^2 &= -f(r)dt^2 + \frac{dr^2}{f(r)} + r^2 d\Omega^2_2~, \\
	f(r) &= 1 - \frac{2m}{r} + \frac{q^2 + p^2}{r^2}~. \\
\end{aligned}\end{equation}
To avoid naked singularities, the charge-to-mass ratio of the black hole must satisfy $\sqrt{q^2 + p^2}/m \leq 1$, with equality corresponding to an extremal dyonic black hole.

Of course, our black hole solution will be modified when we incorporate the higher-derivative operators in equation (\ref{eq:l4}).  These corrections have been worked out perturbatively in the parameters $c_i$ for electric black holes in~\cite{Kats:2006xp,Cheung:2018cwt}, and the extension for dyonic black holes is done in~\cite{Cheung:2019cwi}.  The extremality bound is correspondingly modified to
\begin{equation}
	\frac{\sqrt{q^2 + p^2}}{m} \leq 1 + \frac{2\kappa^2}{5m^2} \Delta + \mathcal{O}\left(\frac{\kappa^4}{m^4}\right)~,
\label{eq:ext}
\end{equation}
where $\Delta$ is given in terms of $c_i$ and the charges as
\begin{equation}
	\Delta = c_1 + c_3 \frac{3p^4 + 4 p^2 q^2 + q^4}{(p^2 + q^2)^2} + 4 c_4 \frac{(p^2 - q^2)^2}{(p^2 + q^2)^2}~.
\label{eq:delta}
\end{equation}
Note that $c_2$ does not appear in $\Delta$, as it is the coefficient of the Gauss-Bonnet term, which is topological and does not affect the equations of motion.  We have dropped higher-order terms in (\ref{eq:ext}) that come from going to higher powers in the perturbation series.    We can justify this truncation by focusing on sufficiently massive black holes such that $m \gg \kappa$.  Moreover, the six-derivative (or higher) operators in our EFT will also contribute to the extremality bound at $\mathcal{O}(\kappa^4/m^4)$ or higher, and so we can also safely ignore their effects in this large black hole limit.

If $\Delta > 0$, then our EFT allows for super-extremal black holes, thus satisfying the requirements for the mild form of the WGC.  Moreover, the extremality correction scales inversely with the black hole mass and approaches zero as $m \to \infty$.  This means that sufficiently massive dyonic black holes (for which (\ref{eq:ext}) is valid) will always be able to decay into smaller black holes with a higher charge-to-mass ratio.  The original motivation for the WGC is based on avoiding black hole remnants by allowing extremal charged black holes to decay, and this is precisely what we recover in this EFT setup.

\section{\label{sec:qm} One-Loop Quantum Corrections} 

We have so far only discussed classical features of black holes of our low-energy effective field theory.  Our goal now is to investigate quantum corrections to these black holes.  These can be systematically analyzed with the background field method by picking a classical saddle-point for our theory and then expanding the fields around their background values via the variations $\delta g_{\mu\nu} = h_{\mu\nu}$, $\delta A_\mu = a_\mu$; the graviton and photon fluctuations $h_{\mu\nu}$ and $a_\mu$ are treated as dynamical fields that propagate on the classical background.  The Lagrangian (\ref{eq:l2}) can then be expanded as $\mathcal{L} = \mathcal{L}_0 + \mathcal{L}_1 + \mathcal{L}_2 + \ldots$, where $\mathcal{L}_n$ is the part of the expansion that contains terms with $n$ powers of the quantum fields.  The leading order piece $\mathcal{L}_0$ contains no field fluctuations and is simply the classical action, evaluated on the background we have chosen, while $\mathcal{L}_1$ is proportional to the equations of motion and will hence vanish.  

\begin{figure}[h]\centering
	\includegraphics[scale = 0.6]{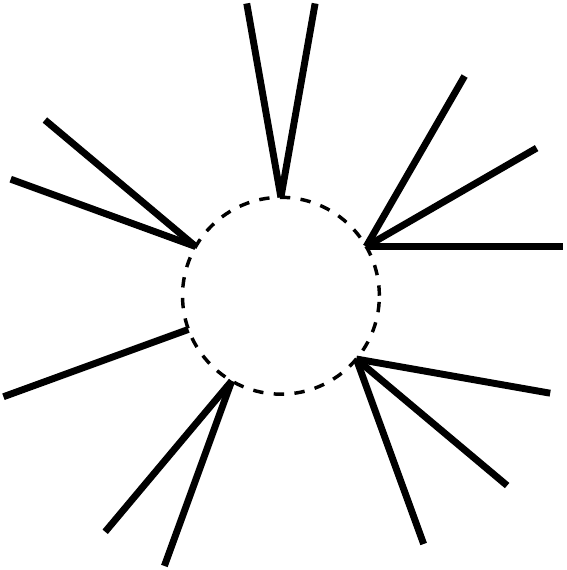}
	\caption{A sample one-loop diagram in the background field method.  The solid lines indicate background fields, while the dotted lines indicate the dynamical field fluctuations around these background fields.  The relevant vertices are all quadratic in the field fluctuations; higher-order interactions will only be relevant for higher-loop diagrams.}
	\label{fig:oneloop}
\end{figure}

It is $\mathcal{L}_2$, the quadratic part of the action, that determines the leading-order quantum corrections to our theory.  To see this, consider one-loop diagrams of the form depicted in figure~\ref{fig:oneloop}, where the vertices involve two internal lines (which represent the dynamical quantum fields) and an arbtirary number of external legs (which represent background fields).  These interactions come from the quadratic Lagrangian $\mathcal{L}_2$, and so the higher-order terms in the expansion of the Lagrangian will not matter for computing one-loop quantum corrections to our classical theory.

Note also that the field fluctuations $a_\mu$ and $h_{\mu\nu}$ should be thought of as gauge fields with associated transformations that leave the quadratic action invariant.  We will use the Faddeev-Popov procedure to pick a particular gauge orbit in the path integral by introducing ghost fields with anti-commuting statistics.  In particular, gauge-fixing the $U(1)$ gauge symmetry of the photon introduces two scalar ghosts $b$ and $c$, while gauge-fixing the diffeomorphism symmetry of the graviton introduces two vector ghosts $b_\mu$ and $c_\mu$.    These ghost fields will appear in the quadratic action, and so we must allow for them to run in the loop as well.

The quadratic action will take the form	$\mathcal{L}_2 = \vec{\phi}\,{}^T \cdot \square \cdot \vec{\phi}$, where $\vec{\phi} = (a_\mu,h_{\mu\nu},b,c,b_\mu,c_\mu)$ is a vector of the field fluctuations and ghosts and $\square$ is a matrix of second-order Laplace-type differential operators that is built out of the background fields and covariant derivatives on the background.  The UV-divergences that arise in one-loop diagrams when the field fluctuations run over the loop with arbitrarily high energies are intimately related to the spectrum of the operator $\square$.  Crucially, these diagrams all exhibit the same UV-divergences, no matter how many external lines are tacked onto the loop~\cite{tHooft:1974toh}.  Using dimensional regularization in $d = 4+\epsilon$ dimensions, the appropriate counterterm Lagrangian that cancels these unphysical divergences is
\begin{equation}
	\mathcal{L}_\text{c.t.} = \frac{\mu^\epsilon a_4(x)}{\epsilon}~,
\label{eq:ct}
\end{equation}
where we have introduced an explicit renormalization scale $\mu$ in order to keep the action dimensionless, and $a_4(x)$ is the $n=2$ case of the heat kernel coefficients $a_{2n}(x)$ of the operator $\square$. These heat kernel coefficients (also known as Seeley-DeWitt coefficients~\cite{DeWitt:1965jb}) are related to the trace of the heat kernel $e^{s\,\square}$ by
\begin{equation}
	\text{tr}\,e^{s\,\square} = \int d^4x\,\sqrt{g}\,\sum_{n=0}^\infty s^{n-2} a_{2n}(x)~.
\end{equation}
The coefficient $a_4(x)$ is also precisely the one-loop contribution to the Weyl anomaly $\langle T\ind{_\mu^\mu}\rangle$.

To compute $a_4(x)$, we will follow the covariant heat kernel method reviewed in~\cite{Vassilevich:2003xt} by bringing the operator $\square$ to the form
\begin{equation}
	\square = (\mathcal{D}_\mu \mathcal{D}^\mu) I + E~,
\end{equation}
where $I$ and $E$ are matrices (acting on the vector $\vec{\phi}$ of field fluctuations) constructed from the background fields and $\mathcal{D}_\mu$ is a matrix of covariant derivatives with an associated two-form field strength $\Omega_{\mu\nu} \equiv [\mathcal{D}_\mu, \mathcal{D}_\nu]$.  The second heat kernel coefficient is then computable by the simple formula~\citep{Vassilevich:2003xt}
\begin{equation}\begin{aligned}
	(4\pi)^2 a_4(x) &= \frac{1}{2}\text{tr}\,E^2 + \frac{1}{12}\text{tr}\,\Omega_{\mu\nu}\Omega^{\mu\nu} \\
	&\quad +\frac{1}{180}(R_{\mu\nu\rho\sigma}R^{\mu\nu\rho\sigma} - R_{\mu\nu}R^{\mu\nu}) \text{tr}\,I~,
\end{aligned}\end{equation}
where we have dropped terms proportional to the Ricci scalar that vanish when evaluated on any Einstein-Maxwell background.  $a_4(x)$ is in general a linear combination of all four-derivative operators in the theory consistent with the symmetries of the low-energy effective field theory, and so the counterterm (\ref{eq:ct}) changes the values of the bare coupling constants in equation (\ref{eq:l4}).

The one-loop quadratic action and the associated heat kernel coefficient $a_4(x)$ on arbitrary Einstein-Maxwell backgrounds has been computed previously in~\cite{Deser:1974cz,Bhattacharyya:2012wz,Charles:2015eha}, the result of which is
\begin{equation}
	a_4(x) = \frac{c}{16\pi ^2} W_{\mu\nu\rho\sigma}W^{\mu\nu\rho\sigma} - \frac{a}{16\pi ^2} E_4~,
\label{eq:a4}
\end{equation}
with $c = \frac{137}{60}$ and $a = \frac{53}{45}$.  All explicit dependence on the field strength cancels out, leaving us with manifestly electromagnetic duality-invariant counterterms.  This is no accident; as shown in~\cite{Charles:2017dbr} this is required by the electromagnetic duality symmetry of the leading-order Einstein-Maxwell theory (\ref{eq:l2}), even if the background itself breaks this symmetry.  This means that the parameters $c_3$ and $c_4$ in the four-derivative Lagrangian (\ref{eq:l4}) are not renormalized at the one-loop level, but $c_1$ and $c_2$ are.

This renormalization of the dimensionless parameters $c_1$ and $c_2$ means that these coefficients are not fixed; they will instead have an associated RG flow and will run with the associated energy scale $\mu$ at which we probe quantum fields propagating on our classical background.  From the counterterm (\ref{eq:a4}), it is straightforward to compute the associated beta functions for $c_1$ and $c_2$ and show that they run logarithmically~\cite{Tong:2014era}.  In particular, we find that
\begin{equation}\begin{aligned}
	c_1(\mu) = c_1(\Lambda) + \frac{c}{16\pi^2} \log\frac{\Lambda}{\mu}~, \\
	c_2(\mu) = c_2(\Lambda) - \frac{a}{16\pi^2} \log\frac{\Lambda}{\mu}~,
\label{eq:run}
\end{aligned}\end{equation}
while the coefficients $c_3$ and $c_4$ will not have logarithmic running below the cut-off scale $\Lambda$.   Since $c > 0$, we find that $c_1(\mu)$ will increase logarithmically in the IR of our theory.  No matter what the primordial value $c_1(\Lambda)$ takes, this logarithmic running will eventually dominate and cause $c_1$ to become positive in the deep IR of the theory.

Our goal now is to understand how this RG flow affects our previous analysis of super-extremal black holes.  The key point here is that if we want to treat black holes as good, stable classical saddle points with small quantum fluctuations propagating on them, we have to focus on very large black holes whose curvature is well below the cut-off scale on our EFT, as it is precisely the curvature of the black hole that sets the characteristic energy scale of the quantum fluctuations around the black hole~\cite{Donoghue:1995cz,Burgess:2003jk}.   The size $m$ of the black hole is inversely proportional to its curvature, and so our EFT should be thought of as describing the quantum fluctuations around black holes at a scale $\mu \sim 1/m \ll \Lambda$.  The larger the black hole background we consider, the smaller the associated curvature at the horizon of the black hole, and so the quantum fluctuations sourced by the background will have typically smaller momenta.

The upshot of this is that, for sufficiently large black holes, the logarithmic running (\ref{eq:run}) of the parameter $c_1$ will dominate and thus $c_1$ scales as
\begin{equation}
	c_1 \sim \frac{c}{16\pi^2} \log\left(m \Lambda\right)~,
\end{equation}
where $c$ is the coefficient of the $W_{\mu\nu\rho\sigma}W^{\mu\nu\rho\sigma}$ term in the counterterm (\ref{eq:a4}).  Since $c > 0$ for Einstein-Maxwell theory, we conclude that one-loop quantum effects cause $c_1$ to become positive.

The super-extremality parameter $\Delta$, as defined in (\ref{eq:delta}), depends on the mass and charges of our black hole background as well as the parameters $c_1$, $c_3$, and $c_4$.  Importantly, though, $c_3$ and $c_4$ are not altered by one-loop quantum effects.  So, no matter what the primordial values of $c_1$, $c_3$, and $c_4$ are at the cut-off scale, black holes with a sufficiently large mass will have $c_1 \gg c_3,c_4$, thus leading to $\Delta \approx c_1 > 0$, independent of the charges of the black hole.

The authors of~\cite{Cheung:2018cwt} argue that the primordial value of $\Delta$, e.g. $\Delta$ as a function of the primordial values of the coefficients $c_i$, must be greater than zero via entropic considerations.  This in turn allows for super-extremal black holes.  What we have shown is that, independent of this argument, the theory automatically allows for super-extremal black holes; even if $\Delta$ starts off negative, it will run to a positive value in the deep IR of our EFT, which corresponds to quantum fields propagating on very massive black hole backgrounds.  It is essential that this running is logarithmic; this means that the value of $\Delta$ is always positive for sufficiently large black holes, but it increases slower than $m$.  The maximum charge-to-mass ratio (\ref{eq:ext}) thus decreases as the size of the black hole increases, guaranteeing that large extremal black holes will always be able to decay into smaller black holes with a larger charge-to-mass ratio.

One might worry that higher-loop quantum effects could spoil this result, but this is fortunately not the case~\cite{Sen:2012dw}.  The two-loop divergences in our theory will renormalize the coefficients of six-derivative EFT operators, such as $R_{\mu\nu}R^{\nu\rho}R\ind{_\rho^\mu}$, $(F_{\mu\nu}F^{\mu\nu})^3$, etc.  By dimensional analysis, these six-derivative operators will contribute to the charge-to-mass bound (\ref{eq:ext}) at $\mathcal{O}(\kappa^4/m^4)$, which is subleading for black holes with $m \gg \kappa$.  Moreover, the effects from even higher-loop diagrams will be even more subleading, and so we can safely ignore the effects from our massless fields running in higher-loop diagrams.

The results so far have been for a theory whose massless field content comprises only a graviton and a photon.  We will now investigate how these results change when we add additional massless fields to the theory that can run in loops.  In particular, we will consider adding minimally-coupled neutral particles with spin $s \leq 3/2$ and see how they affect the running of the coefficients $c_i$ in the four-derivative Lagrangian (\ref{eq:l4}).

We still want to look at quantum corrections to Einstein-Maxwell solutions (and in particular dyonic black holes), and so we will turn off all of these minimally-coupled fields in our classical background.  When we expand our action around this classical saddle point, though, these additional fields will also fluctuate around the background, and so we must keep track of their contribution to the one-loop quadratic action.  Since they are neutral and minimally-coupled, their corresponding heat kernel coefficients will not have any explicit dependence on the background field strength, and thus they will also take the form (\ref{eq:a4}), with the values of $c$ and $a$ depending on the spin of the particle.  These coefficients are well-known~\cite{
Christensen:1978md,Birrell:1982ix}, and the results are listed in table~\ref{tab:ca}.

\bgroup
\def\arraystretch{1.5}
\begin{table}[ht]
\centering
\begin{tabular}{|c|c|c|}\hline
{Spin} & $c$& $a$ \\ \hline
$0$ & $\frac{1}{120}$ & $\frac{1}{360}$ \\ \hline
$1/2$ & $\frac{1}{20}$ & $\frac{11}{360}$ \\ \hline
$1$ & $\frac{1}{10}$ & $\frac{31}{180}$ \\ \hline
$3/2$ & $-\frac{77}{60}$ & $-\frac{229}{360}$ \\ \hline
\end{tabular}
\caption{\label{tab:ca} Contributions to $c$ and $a$ from minimally-coupled fields of different spins.  The fermions are assumed to be Dirac fermions, but we can obtain the result for Weyl or Majorana fermions simply by dividing by two.}
\end{table}

What we can clearly see is that scalars, fermions, and vector fields all contribute positively to $c$, which in turn means that these fields will enhance the positive running of $c_1$ in the IR.  If we want to stabilize things such that $c_1$ does not run at one-loop, our only option is to include gravitino fields into our theory.  The simplest option would be to include a single gravitino, but this is not enough; the result for Einstein-Maxwell theory with a free gravitino is $c = \frac{137}{60} - \frac{77}{60} = 1$, up to the inclusion of additional matter with spin $s \leq 1$, and thus we would still have positive running in the deep IR.  

This indicates that, if we want an EFT in which super-extremal black holes are not allowed, we need to include multiple gravitino fields, which naturally points us towards $\mathcal{N} \geq 2$ supergravity theories.  Of course, the values of $c$ and $a$ given in table~\ref{tab:ca} are only for minimally-coupled fields, and supersymmetry will require specific couplings between fields.  In the next section, we will study how to embed Einstein-Maxwell solutions into $\mathcal{N} \geq 2 $ supergravity and analyze the contributions to $c$ and $a$ from all fields in the theory when these supersymmetric couplings are accounted for.

\section{\label{sec:n2} \texorpdfstring{$\mathcal{N} \geq 2$ Supergravity}{N >= 2 Supergravity}}

The leading-order action (e.g. containing only relevant operators with two derivatives) for the massless bosonic fields of a four-dimensional ungauged $\mathcal{N}=2$ supergravity theory with $n_V$ vector multiplets is
\begin{equation}
	\mathcal{L} = \frac{1}{2\kappa^2}R - g_{\alpha \bar{\beta}} \nabla_\mu z^\alpha \nabla^\mu \bar{z}^{\bar{\beta}} + \frac{1}{2}\text{Im}\left(\mathcal{N}_{IJ} F^{+I}_{\mu\nu} F^{+J\mu\nu}\right)~,
\label{eq:l2susy}
\end{equation}
where the indices $\alpha = 1,\ldots,n_V$ enumerate the vector multiplets, while $I = 0,\ldots,n_V$ enumerate the vector fields in the vector multiplets as well as the graviphoton field in the gravity multiplet.  Note also that $F^{\pm I}_{\mu\nu}$ denotes the (anti)-self-dual part of the field strength $F^I_{\mu\nu}$.  The interactions between the vector fields are determined by the prepotential $F(X)$, a holomorphic function of the projective coordinates $X^I$ on the K\"ahler manifold.  If we denote derivatives of the prepotential with respect to these coordinates by $F_I  = \frac{\partial F(X)}{\partial X^I}$, the matrix $\mathcal{N}_{IJ}$ is given by
\begin{equation}
	\mathcal{N}_{IJ} = \bar{F}_{IJ} +2 i \frac{\text{Im}\,F_{IK} X^K \text{Im}\,F_{JL}X^L}{\text{Im}\,F_{MN} X^M X^N}~,
\end{equation}
where $\bar{F}$ denotes the anti-holomorphic complex conjugate of $F$.  The projective coordinates are in turn specified as a function of the physical scalars $z^\alpha$.  The K\"ahler metric $g_{\alpha \bar{\beta}}$ that determines the scalar kinetic term is given by
\begin{equation}
	g_{\alpha \bar{\beta}} = i \frac{\partial}{\partial z^{\alpha}} \frac{\partial}{\partial \bar{z}^{\bar{\beta}}}\left(\bar{F}_I X^I - F_I \bar{X}^I\right)~.
\end{equation}
Note that the normalization of these projective coordinates is chosen such that $\text{Im}\,F_{IJ} X^I \bar{X}^J = -1/2\kappa^2$.

More general (and in particular multi-centered) black holes can be explicitly constructed in supergravity, but these require giving the scalar fields non-trivial profiles according to the precise form of the prepotential.  The moduli space for these more general solutions can contain marginal walls of stability~\cite{Denef:2000nb,Denef:2007vg}, which can allow extremal black holes to decay without requiring a super-extremal decay product state.  The WGC is thus much less well-motivated for these situations.  We will therefore restrict our attention to single-centered, dyonic Reissner-Nordstr\"om black holes and try to understand what becomes of our previous WGC analysis when we consider them in an $\mathcal{N} = 2$ supergravity setting.

There is a simple prescription to embed arbitrary solutions to Einstein-Maxwell theory into $\mathcal{N} = 2$ supergravity in a prepotential-independent manner.  We first turn off all fermions in the background.  Then, we choose the scalars $z^\alpha$ to be some arbitrary constants, which in turn fixes the projective coordinates $X^I$ to some constant values.  We then choose all field strengths to be given in terms of the Einstein-Maxwell field strength $F_{\mu\nu}$ and the projective coordinates as
\begin{equation}
	F^{+I}_{\mu\nu} = X^I F_{\mu\nu}^+~.
\label{eq:align}
\end{equation}
As shown in~\cite{Charles:2015eha}, this embedding automatically satisfies all equations of motion, with an energy-momentum tensor that is identical to that of Einstein-Maxwell theory, and so the resulting geometry is unchanged.  In essence, we are just identifying the $U(1)$ gauge field that the black hole is charged under with the $U(1)_R$ gauge field in the $\mathcal{N}=2$ gravity multiplet.  We then demand that all vector multiplet gauge fields are also proportional to this such that the background geometry is unchanged.

We can additionally add in matter in the form of hyper multiplets in a trivial way: the scalars in the hyper multiplet are minimally coupled to the background, and we want to turn off all fermions in the background, and so we are free to simply set all hyper multiplet fields to zero on our background.

We can also easily extend the embedding to $\mathcal{N} > 2$ supergravity by decomposing everything into $\mathcal{N}=2$ multiplets.  Once we have designated one of the gravity multiplet gauge fields to be the $U(1)_R$ graviphoton field and decompose accordingly, we are left with (in $\mathcal{N}=2$ language) a single gravity multiplet, $\mathcal{N}-2$ gravitino multiplets, $n_V$ vector multiplets, and $n_H$ hyper multiplets, where $n_V$ and $n_H$ depend on the number of additional matter multiplets we consider in the $\mathcal{N} > 2$ supergravity theory.  The gauge fields in the gravitino multiplets are minimally-coupled to the background, and so we can consistently turn all gravitino multiplet fields off in the background, thus preserving our Einstein-Maxwell embedding for $\mathcal{N} > 2$ supergravity as well.

Classically, the charge-to-mass ratio of Reissner-Nordstr\"om black holes in our $\mathcal{N}\geq 2$ embedding is the same bound as in Einstein-Maxwell theory, so $\sqrt{q^2 + p^2}/m \leq 1$.  In order to ask how the charge-to-mass ratio is modified by higher-derivative terms in the action, we first need to ask what kinds of higher-derivative terms are allowed.  Supersymmetry is very constraining in this regard, and the off-shell $\mathcal{N}=2$ superconformal calculus approach can be used to write down two independent operators that can appear at four-derivative order; these should be thought of as the supersymmetric completions of $W_{\mu\nu\rho\sigma}W^{\mu\nu\rho\sigma}$~\cite{LopesCardoso:2000qm} and of the Gauss-Bonnet invariant~\cite{Butter:2013lta}, and they take the schematic form
\begin{equation}\begin{aligned}
	\mathcal{O}_1 &= W_{\mu\nu\rho\sigma}W^{\mu\nu\rho\sigma} + \text{(SUSY matter)}~, \\
	\mathcal{O}_2 &= E_4 + \text{(SUSY matter)}~.
\label{eq:o2}
\end{aligned}\end{equation}
We could in principle also try to include operators of the (schematic) form $RF^2$ and $F^4$, but these are heavily constrained by the $U(1)_R$ symmetry and symplectic invariance of the action.  When these constraints are combined with the alignment (\ref{eq:align}) of all field strengths, the result is that such operators will reduce purely to geometric invariants~\cite{Charles:2017dbr}, and thus the basis of four-derivative operators (\ref{eq:o2}) is complete within the scope of our Einstein-Maxwell embedding.

So, the four-derivative Lagrangian in our $\mathcal{N}\geq 2 $ supergravity EFT will be a linear combination of the two operators (\ref{eq:o2}).  For the purposes of determining the leading-order corrections to the background coming from this four-derivative Lagrangian, we are free to impose the equations of motion coming from the two-derivative Lagrangian (\ref{eq:l2susy}).  When we do so, though, something interesting happens: both simply become (up to irrelevant total covariant derivatives) the Gauss-Bonnet invariant~\cite{Charles:2017dbr}.    This means that, for arbitrary Einstein-Maxwell solutions to $\mathcal{N}=2$ supergravity, the only four-derivative term consistent with supersymmetry that we can write down in our EFT is $E_4$, which corresponds to demanding that the coefficients $c_1 = c_3 = c_4 = 0$ in (\ref{eq:l4}).  This in turn sets the tree-level value of $\Delta$, as defined in (\ref{eq:delta}), to be $\Delta = 0$ for dyonic Reissner-N\"ordstrom black holes in $\mathcal{N} \geq 2$ supergravity.

\bgroup
\def\arraystretch{1.5}
\begin{table}[ht]
\centering
\begin{tabular}{|c|c|c|c|}\hline
\multicolumn{2}{|c|}{\textbf{Multiplet}} & $c$& $a$ \\ \hline \hline
\multirow{3}{*}{Hyper} & Bosons & $\frac{1}{30}$ & $\frac{1}{90}$ \\ \cline{2-4}
& Fermions & $-\frac{1}{30}$ & $-\frac{19}{360}$ \\ \cline{2-4}
& Total & $0$ & $-\frac{1}{24}$ \\ \hline \hline
\multirow{3}{*}{Vector} & Bosons & $-\frac{1}{20}$ & $\frac{1}{90}$ \\ \cline{2-4}
& Fermions & $\frac{1}{20}$ & $\frac{11}{360}$ \\ \cline{2-4}
& Total & $0$ & $\frac{1}{24}$ \\ \hline \hline
\multirow{3}{*}{Gravitino} & Bosons & $\frac{1}{5}$ & $\frac{31}{90}$ \\ \cline{2-4}
& Fermions & $-\frac{1}{5}$ & $\frac{41}{360}$ \\ \cline{2-4}
& Total & $0$ & $\frac{11}{24}$ \\ \hline \hline
\multirow{3}{*}{Gravity} & Bosons & $\frac{137}{60}$ & $\frac{106}{90}$ \\ \cline{2-4}
& Fermions & $-\frac{137}{60}$ & $-\frac{589}{360}$ \\ \cline{2-4}
& Total & $0$ & $-\frac{11}{24}$ \\ \hline
\end{tabular}
\caption{\label{tab:can2} The values of $c$ and $a$ for each massless $\mathcal{N}=2$ multiplet when we embed Einstein-Maxwell theory into $\mathcal{N}\geq 2$ supergravity and compute the corresponding heat kernel coefficients.  Each multiplet is defined such that it has $4+4$ bosonic and fermionic degrees of freedom, so e.g. the hyper multiplets are complex.}
\end{table}

Our analysis so far is purely classical.  We now need to look at quantum corrections and see if these can shift the classical values of $c_i$.  That is, we need to expand all fields in the theory around their background values, compute the one-loop quadratic action for all fields, and from this extract the heat kernel coefficient $a_4(x)$ that controls divergences.  This procedure has been done before in~\cite{Charles:2015eha}.  Importantly, even if the background is non-supersymmetric, the fluctuations of the fields around the Einstein-Maxwell background respect supersymmetry and can be organized into $\mathcal{N}=2$ multiplets that decouple from one another.  Explicit computation shows that the resulting heat kernel coefficients have no explicit dependence on the background field strength, and so they can all be written in the form (\ref{eq:a4}), where the values of $c$ and $a$ are dependent on which multiplet we look at.  The results of this computation are tabulated in table~\ref{tab:can2}.

The main takeway from table~\ref{tab:can2} is that the fermions and bosons in each multiplet conspire to have exactly opposite contributions to $c$, and so the result is that $c = 0$ for any full $\mathcal{N} = 2$ multiplet.  This cancellation hinges upon the fact that the fields are not simply minimally-coupled; instead, $\mathcal{N}=2$ supersymmetry demands that the hyper fermions, the vector multiplet bosons, and the gravitini all couple in a non-trivial way to the background field strength.  These non-minimal couplings are such that their contribution to $c$ is negative, and in fact exactly cancels their corresponding superpartners.  We are left with
\begin{equation}
	a_4(x) = \frac{1}{384\pi^2}\left(11 - 11 (\mathcal{N}-2) - n_V + n_H\right) E_4~,
\end{equation}
and thus only the coefficient of $E_4$ in the action is renormalized at one-loop level.  The classical constraint $c_1 = c_3 = c_4 = 0$ on the four-derivative terms in the action is thus preserved by one-loop quantum effects.  This in turn means that $\Delta = 0$ holds at the one-loop quantum level as well.  We therefore conclude that dyonic Reissner-Nordstr\"om black holes in $\mathcal{N} \geq 2$ supergravity theories are not allowed to be super-extremal, even when one-loop quantum effects are accounted for.

Of course, any massive state charged under the $U(1)_R$ gauge field in $\mathcal{N}=2$ supergravity must obey the BPS bound
\begin{equation}
	\frac{\sqrt{q^2 + p^2}}{m} \leq 1~,
\end{equation}
which arises from demanding at the level of the supersymmetry algebra that such massive states have non-negative norm.  Moreover, the BPS bound is protected from perturbative quantum effects (and even from non-perturbative effects), which automatically demands that one-loop quantum effects cannot generate a non-zero value for $\Delta$.  However, it is important to note that our analysis holds for arbitrary Einstein-Maxwell solutions embedded into supergravity, including ones that break supersymmetry.  This means we have shown something stronger: even if the black hole is non-BPS, it is still protected from one-loop corrections.

This foray into $\mathcal{N} \geq 2$ supergravity also gives us insight into what kind of couplings are required in order to shift the value of $c$ to be negative.  For example, the hyper multiplets each contain two massless Majorana fermions that are coupled to one another via a Pauli term.  The corresponding one-loop quadratic Lagrangian for these hyper fermions takes the form
\begin{equation}
	\mathcal{L} = - \bar{\psi}^A \gamma^\mu \nabla_\mu \psi^A + \lambda \bar{\psi}^A F_{\mu\nu}\gamma^{\mu\nu}\psi^B \varepsilon_{AB} ~,
\label{eq:maj}
\end{equation}
for some numerical coupling constant $\lambda$.  In particular, $\lambda = 1/4$ for the hyper fermion fluctuations around Einstein-Maxwell backgrounds, but we will leave $\lambda$ general for now.  From here, we can go through the covariant heat kernel method and compute the heat kernel coefficient $a_4(x)$.  Despite the explicit appearance of $F_{\mu\nu}$ in (\ref{eq:maj}), computation shows that $a_4(x)$ still takes the form of (\ref{eq:a4}), with
\begin{equation}\begin{aligned}
	c &= \frac{1}{20} - \frac{8 \lambda^2}{3} + \frac{64 \lambda^4}{3}~, \\
	a &= \frac{11}{360} - \frac{8 \lambda^2}{3} + \frac{64 \lambda^4}{3}~.
\end{aligned}\end{equation}
The value of $c$ is positive for all $\lambda$ except on a very small range of values between $|\lambda| \approx 0.15$ and $|\lambda| \approx 0.32$.  In particular, $c$ is minimized at $\lambda = 1/4$, which is precisely the value dictated by supersymmetry.  This means that it is not enough to simply have Pauli couplings; obtaining a negative value of $c$ requires a good deal of fine-tuning on the strength of the Pauli couplings, a fine-tuning that supersymmetry implements automatically.  Nonetheless, it is conceivably possible to engineer a theory in which enough finely-tuned matter is added to force $c < 0$, which would in turn prevent super-extremal black hole states.

\section{\label{sec:con} Conclusions}

\begin{figure}[ht]\centering
	\includegraphics[scale = 0.8]{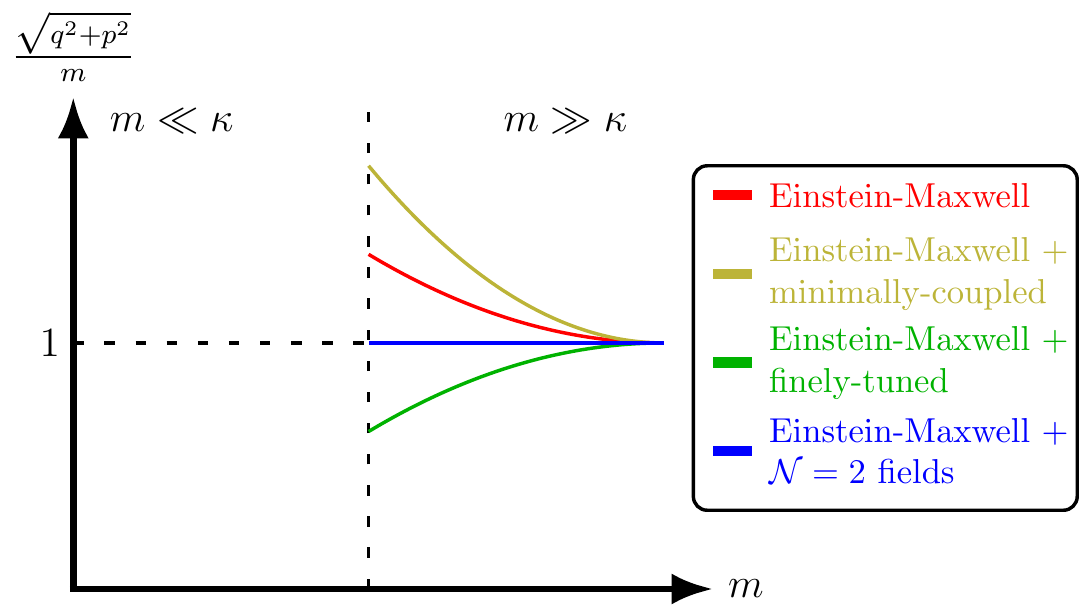}
	\caption{The maximum charge-to-mass ratio for various Einstein-Maxwell EFTs with additional massless matter fields.  The one-loop quantum effects dominate for very large black holes, but for smaller black holes such an EFT analysis is insufficient (as indicated by the dotted vertical line).  As $m\to\infty$, the bound approaches unity.}
	\label{fig:qmplot}
\end{figure}

In this paper, we have shown that one-loop quantum effects in a large class of low-energy EFTs with a graviton and a photon automatically allow for dyonic super-extremal black holes, independent of any assumptions being made on the UV completion of the theory.  Additionally, we have shown that this is not allowed when we consider Einstein-Maxwell as a subsector of a full $\mathcal{N} \geq 2 $ supergravity theory; supersymmetry dicates couplings to the background field strength that are precisely the ones required to enforce the BPS bound.  Moreover, if we want the charge-to-mass ratio bound to be less than one, it requires additional matter with very finely-tuned couplings.  These results are summarized in figure~\ref{fig:qmplot}.  The main conclusion we can draw is that the WGC is almost universal, in the sense that it takes a lot of work to engineer a theory in which super-extremal black holes do not automatically arise from quantum effects.

We can also view our results as evidence for an intriguing relationship between black hole entropy and the WGC.  In particular, the one-loop running of the EFT coefficients (\ref{eq:run}) that resulted in super-extremal black holes also yields a logarithmic correction to the Bekenstein-Hawking area law~\cite{Bhattacharyya:2012wz,Sen:2012dw}.  This logarithmic term places stringent constraints on any microscopic description of black hole entropy.  Previous studies~\cite{Fisher:2017dbc,Cottrell:2016bty} also hint at a deep relationship between this logarithmic correction and the WGC, and we look forward to exploring this further.  

Another perspective one could take is that the mild version of the WGC, where we require at least one super-extremal state in the theory, is too easy to satisfy and hence isn't a good tool for mapping out the swampland.  This is a viewpoint echoed in myriad recent examples~\cite{Heidenreich:2016aqi,Montero:2016tif,Aalsma:2019ryi,Heidenreich:2019zkl} that demonstrate a stronger version of the WGC with an infinite tower of super-extremal states.  It is interesting to note that many of these stronger forms of the WGC are intricately tied to modular invariance, which is also a key component in precision black hole microstate counting.  We will leave more rampant speculation on this subject to future work, though.

The most natural extension of this work is to explore how the charge-to-mass ratio of more exotic black holes are altered by one-loop quantum effects.  One particularly nice example would be black holes in an Einstein-Maxwell-Dilaton EFT with a non-trivial dilaton profile turned on, which arise naturally in the low-energy limit of heterotic string theory~\cite{Garfinkle:1990qj}.  It would also be worthwhile to adapt our methods to asymptotically-AdS black holes in order to probe the holographic WGC~\cite{Nakayama:2015hga,Montero:2018fns}.

\section*{Acknowledgments}

We thank Nikolay Bobev, Miguel Montero, and Thomas Van Riet for useful discussions and feedback, as well as Finn Larsen for collaboration on earlier related work.  AMC is supported in part by the KU Leuven C1 grant ZKD1118 C16/16/005, the National Science Foundation of Belgium (FWO) grant G.001.12 Odysseus, and by the European Research Council grant no. ERC-2013-CoG 616732 HoloQosmos.

\bibliographystyle{apsrev4-1}
\bibliography{wgclogs}

\end{document}